\RequirePackage{ifpdf}

\documentclass{PoS}
\pdfoutput=1
\usepackage{graphicx,times, amsmath, epsfig}

\newcommand{\Ts}{T_{\rm S}}
\newcommand{\Tk}{T_{\rm K}}

\newcommand{\nf}{x_{\rm HI}}
\newcommand{\avenf}{\bar{x}_{\rm HI}}

\newcommand{\lya}{Ly$\alpha$}

\newcommand{\Msun}{M_\odot}
\newcommand{\Tvir}{T_{\rm vir}}

\newcommand{\Tcmb}{T_\gamma}
\newcommand{\delT}{\delta T_b}

\newcommand{\delNL}{\delta_{\rm nl}}
\newcommand{\Mmin}{M_{\rm min}}

\newcommand\lsim{\mathrel{\rlap{\lower4pt\hbox{\hskip1pt$\sim$}}
        \raise1pt\hbox{$<$}}}
\newcommand\gsim{\mathrel{\rlap{\lower4pt\hbox{\hskip1pt$\sim$}}
        \raise1pt\hbox{$>$}}}
\def\myputfigure#1#2#3#4#5%
{\vskip#5pt\makebox[0pt]{\hskip#2in
\includegraphics[width=#3\textwidth]{#1}}\vskip#4pt\hfill}

\newenvironment{packed_enum}{
\begin{enumerate}
  \setlength{\itemsep}{1pt}
  \setlength{\parskip}{0pt}
  \setlength{\parsep}{0pt}
}{\end{enumerate}}

\newenvironment{packed_item}{
\begin{itemize}
  \setlength{\itemsep}{1pt}
  \setlength{\parskip}{0pt}
  \setlength{\parsep}{0pt}
}{\end{itemize}}

\title{Constraining the Astrophysics of the Cosmic Dawn and the Epoch of Reionization with the SKA}

\ShortTitle{Constraining CD and EoR astrophysics with the SKA}

\author{
\speaker{Andrei Mesinger}$^1$, 
Andrea Ferrara$^1$,
Bradley Greig$^1$,
Ilian Iliev$^2$,
Garrelt Mellema$^3$,
Jonathan Pritchard$^4$,
Mario G. Santos$^{5,6,7}$,
on behalf of the EoR/CD Science Team
\\ 
$^1$Scuola Normale Superiore, Piazza dei Cavalieri, 7  56126 Pisa, Italy;
$^2$Astronomy Centre, Department of Physics and Astronomy, University of Sussex, Falmer, Brighton BN1 9QH, UK;
$^3$Dept. of Astronomy \& Oskar Klein Centre, Stockholm University, AlbaNova, SE-10691 Stockholm, Sweden;
$^4$Department of Physics, Blackett Laboratory, Imperial College, London SW7 2AZ, UK;
$^5$Department of Physics, University of Western Cape, Cape Town 7535, South Africa;
$^6$SKA SA, 4rd Floor, The Park, Park Road, Pinelands, 7405, South Africa;
$^7$CENTRA, Departamento de Fisica, Instituto Superior Tecnico, 1049-001 Lisboa, Portugal
\\
E-mail: \email{andrei.mesinger at sns.it}
}

\abstract{

The Square Kilometre Array (SKA) will offer an unprecedented view onto the early Universe, using interferometric observations of the redshifted 21cm line. The 21cm line probes the thermal and ionization state of the cosmic gas, which is governed by the birth and evolution of the first structures in our Universe. Here we show how the evolution of the 21cm signal will allow us to study when the first generations of galaxies appeared, what were their properties, and what was the structure of the intergalactic medium.  We highlight qualitative trends which will offer robust insights into the early Universe.

}

\FullConference{
Advancing Astrophysics with the Square Kilometre Array\\
June 8-13, 2014\\
Giardini Naxos, Italy}

\begin{document}

\section{Introduction}

\begin{figure*}
\vspace{-1\baselineskip}
{
\includegraphics[width=\textwidth]{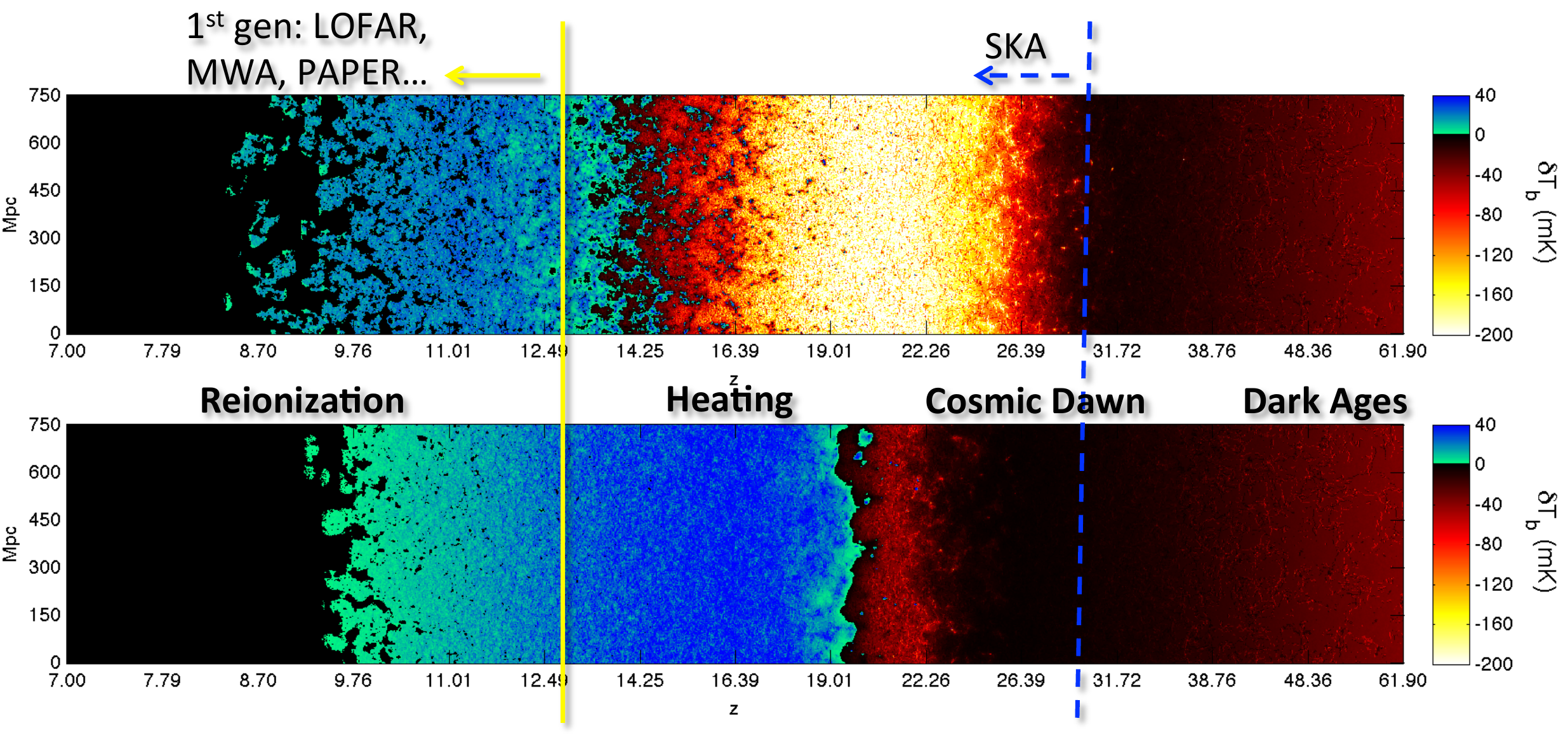}
}
\vspace{-\baselineskip}
\caption{\footnotesize
 21cm brightness temperature offset from the CMB (from \cite{MFS13}).  The horizontal axis shows evolution along the comoving line-of-sight coordinate, from $z\approx62$ to  $z\approx7$. 
 From right to left we see the expected major milestones in the signal: (i) collisional decoupling ({\it red$\rightarrow$black}); (ii) WF coupling ({\it black$\rightarrow$yellow}); (iii) IGM heating ({\it yellow$\rightarrow$blue}); (iv) reionization ({\it blue$\rightarrow$black}).
The {\it top panel} corresponds to a ``fiducial'' model,
 while the {\it lower panel} corresponds to an ``extreme X-ray'' model in which primordial galaxies are much more efficient than local star-bursts  in generating X-rays (see, e.g. \cite{Mirabel11}), saturating the unresolved soft X-ray background \cite{HM07} by $z\sim10$. The  lower panel further assumes that soft X-rays with energies $\lsim$1keV are absorbed within the host galaxy.  Although the models have comparable electron scattering optical depths, $\tau_e$, the astrophysical milestones are very different in 21cm.
The redshift limit accessible to first generation interferometers roughly corresponds to the vertical yellow line; the SKA-LOW should probe out to roughly the vertical blue line, opening-up a new window on the Cosmic Dawn.
The slices are 750 Mpc in height and 1.5 Mpc  thick.
}
\label{fig:delT}
\vspace{-0.5\baselineskip}
\end{figure*}

With unprecedented resolution and sensitivity, the SKA-LOW will enable ground-breaking studies of early Universe astrophysics through the 21cm line from neutral hydrogen.  No other planned instrument will allow us to study baryons at high redshift in such detail.  We will be able to trace the delicate, physics-rich interplay between the intergalactic medium (IGM) and the first galaxies.

As a cosmological probe, the signal is usually represented in terms of the offset of the 21cm brightness temperature from the cosmic microwave background (CMB) temperature, $\Tcmb$, along a line of sight at observed frequency $\nu$ (c.f. \cite{FOB06}):
\begin{align}
  &\delT(\nu) = \frac{\Ts - \Tcmb}{1+z} (1 - e^{-\tau_{\nu_0}}) \approx \\
\nonumber &27 \nf \left(1 - \frac{\Tcmb}{\Ts}\right) \left(1+\delNL\right) \left(\frac{H}{dv_r/dr + H}\right) \sqrt{\frac{1+z}{10} \frac{0.15}{\Omega_{\rm M} h^2}} \left( \frac{\Omega_b h^2}{0.023} \right) \left( \frac{1-Y_p}{0.75} \right) {\rm mK},
\end{align}
\noindent where $T_S$ is the gas spin temperature, $\tau_{\nu_0}$ is the optical depth at the 21cm frequency $\nu_0$, $\delNL({\bf x}, z) \equiv \rho/\bar{\rho} - 1$ is the evolved (Eulerian) density contrast, $Y_p$ is the Helium mass fraction, $H(z)$ is the Hubble parameter, $dv_r/dr$ is the comoving gradient of the line of sight component of the comoving velocity, and all quantities are evaluated at redshift $z=\nu_0/\nu - 1$.  The cosmological 21cm signal uses the CMB as a back-light: if $\Ts < \Tcmb$, then the gas is seen in absorption, while if $\Ts > \Tcmb$, the gas is seen in emission.


The spin temperature, $\Ts$, interpolates between the CMB temperature, $\Tcmb$, and the gas kinetic temperature, $T_K$.  Since we only observe the {\it contrast} of the gas against the CMB, a signal is only obtained if $T_S \rightarrow T_K$.  This coupling is achieved through either: (i) collisions, which are effective in the intergalactic medium (IGM) at high redshifts, $z\gsim50$; or (ii) a Lyman alpha background (so-called Wouthuysen-Field (WF) coupling; \cite{Wouthuysen52, Field58}), effective soon after the first sources turn on.

In the top panel of Fig. \ref{fig:delT}, we show a slice through the $\delT$ field in a ``fiducial'' model (below we use the term ``fiducial'' to refer to models in which atomically-cooled galaxies have similar X-ray and UV properties as local ones, with an ionizing emissivity such that the mid-point of reionization is at $z\sim10$; for further details, see, e.g. \cite{MFS13}).  It is immediately obvious that the 21cm signal is a physics-rich probe, encoding information on various processes during the Cosmic Dawn (CD) and Epoch of Reionization (EoR).  Although the exact timing of the cosmic epochs is uncertain, the relative order is robustly predicted (c.f. \cite{Furlanetto06}; \S 2.1 in \cite{MO12}):
\begin{packed_enum}
\item {\bf Collisional coupling}: The IGM is dense at high redshifts, so the spin temperature is uniformly collisionally coupled to the gas kinetic temperature, $T_K = T_S \lsim \Tcmb$. Following thermal decoupling from the CMB ($z\lsim300$), the IGM cools adiabatically as $T_K \propto (1+z)^2$, faster than the CMB: $\Tcmb \propto (1+z)$.  Thus $\bar{\delT}$ is negative.  This epoch, serving as a {\it clean probe of the matter power spectrum} at $z\gsim100$, is not shown in Fig. \ref{fig:delT}.
\item {\bf Collisional decoupling}:  The IGM becomes less dense as the Universe expands.  The spin temperature starts to decouple from the kinetic temperature, and begins to approach the CMB temperature again, $T_K < T_S \lsim \Tcmb$.  Thus $\bar{\delT}$ starts rising towards zero.  Decoupling from $\Tk$ occurs as a function of the local gas density, with underdense regions decoupling first. Fluctuations are sourced by the density field, and again {\it offer a direct probe of cosmology}. Eventually ($z\sim25$), all of the IGM is decoupled and there is little or no signal.  This epoch corresponds to the red$\rightarrow$black transition on the right edge of Fig. \ref{fig:delT}.
\item {\bf WF coupling (i.e. Ly$\alpha$ pumping)}:  The first astrophysical sources turn on, and begin coupling $T_S$ and $T_K$, this time through the \lya\ background. $\bar{\delT}$ becomes more negative, reaching values as low as $\bar{\delT}\sim$-100 -- -200 mK (depending on the offset of the WF coupling and X-ray heating epochs). Fluctuations are driven by the strength of the \lya\ background. This epoch, offering a window on the {\it very first stars in our Universe}, corresponds to the black$\rightarrow$yellow transition in Fig. \ref{fig:delT}.
\item{\bf IGM heating}: The IGM is heated, with the spin temperature now coupled to the gas temperature, $T_K = T_S$. Fluctuations are sourced by the gas temperature.  As the gas temperature surpasses $\Tcmb$, the 21cm signal changes from absorption to emission, becoming insensitive to the actual value of $T_S$ (see eq. 1).  This epoch probes all processes which heat the IGM, {\it both astrophysical and cosmological}.  The dominant source of heating is likely the X-rays from early accreting black holes (e.g. \cite{Furlanetto06}) or from the hot interstellar medium (ISM; \cite{Pacucci14}).
  This epoch corresponds to the yellow$\rightarrow$blue transition in the panels of Fig. \ref{fig:delT}.
\item {\bf Reionization}:  as the abundance of {\it early galaxies} increases, the IGM gradually becomes ionized, a process which is inside-out on large scales.  Fluctuations during the advanced stages are dominated by the ionization field.  The tomography of this process is sensitive to the nature and clustering of the dominant UV sources (e.g. \cite{McQuinn07}), as well as the evolution of inhomogeneous recombinations (e.g. \cite{SM14}).  The 21cm signal decreases, approaching zero.  This epoch corresponds to the blue$\rightarrow$black transition in the panels of Fig. \ref{fig:delT}.
\end{packed_enum}

The first two stages (the Dark Ages) allow us to probe cosmology at redshifts much lower than recombination, while the last three stages are sensitive to early astrophysical sources (and sinks) of cosmic radiation fields.  These last three stages will be observable with SKA1-LOW.
Below, we focus on the astrophysical insight which can be gained from the CD and EoR (for cosmological insights, see \S Cosmology from the EoR/CD).
%
%
As a foreshadowing of how astrophysics can impact the CD and EoR signals, in the bottom panel of Fig. \ref{fig:delT}, we show an alternate model, in which the early galaxies formed later, but were much more efficient at producing hard X-ray photons, saturating the unresolved X-ray background (XRB; \cite{HM07}) by $z\approx10$; the CD and EoR are dramatically different in these two models.

As our observable, we focus on the 21cm power spectrum.  Although alternate statistics can provide complimentary observations, the power spectrum serves to quantify the main impact of astrophysics on the morphology of the 21cm signal.  We define the 3D power spectrum as $P_{21}(k, z) = k^3/(2\pi^2 V) ~ \bar{\delT}(z)^2 \langle|\delta_{\rm 21}({\bf k}, z)|^2\rangle_k$, where $\delta_{21}({\bf x}, z) \equiv \delT({\bf x}, z)/ \bar{\delT}(z) - 1$. Moreover, when discussing detectability, we focus on the redshift evolution of large-scale power at $k\approx0.2$ Mpc$^{-1}$ (e.g. \cite{Baek10}), roughly corresponding to the largest scales which should be relatively foreground free (e.g. \cite{Pober13}).  Our models of the SKA1-LOW thermal noise are described in \cite{ME-WH14}.
In brief, we take a fiducial 1000h observation, with a $\Delta z = 0.5$ bandwidth and a frequency resolution of 1 kHz. 866 stations (each  a 17x17 array of log-periodic dipoles) are placed using a Gaussian distribution with 75 \% falling within 1000m of the center \cite{Dewdney13}. For most frequency bins and reasonable astrophysical parameters, the S/N can be dominated by cosmic variance (see \cite{ME-WH14} and Fig. \ref{fig:heating}).  This suggest significant results already with a 50\% early-science phase with SKA1-LOW, and also motivates a multi-tiered strategy combining several, moderately deep observations
(as highlighted in the EoR/CD science goals; see also \S \ref{sec:obs}).

Unless stated otherwise, we quote all quantities in comoving units.
The predictions below are consistent with recent Planck measurements of cosmological parameters \cite{Planck13}.


Below we highlight how SKA1-LOW will allow us to study the galaxies and IGM during the CD/EoR.  It will allow us to answer some of the most fundamental questions in astrophysical cosmology: {\it When did the first generations of galaxies appear?  What were their UV and X-ray properties? What was the small-scale structure of the IGM?}


\section{First, molecularly-cooled galaxies}
\label{sec:mol}

\begin{figure*}
\vspace{-1\baselineskip}
\begin{center}
{
\includegraphics[width=0.45\textwidth]{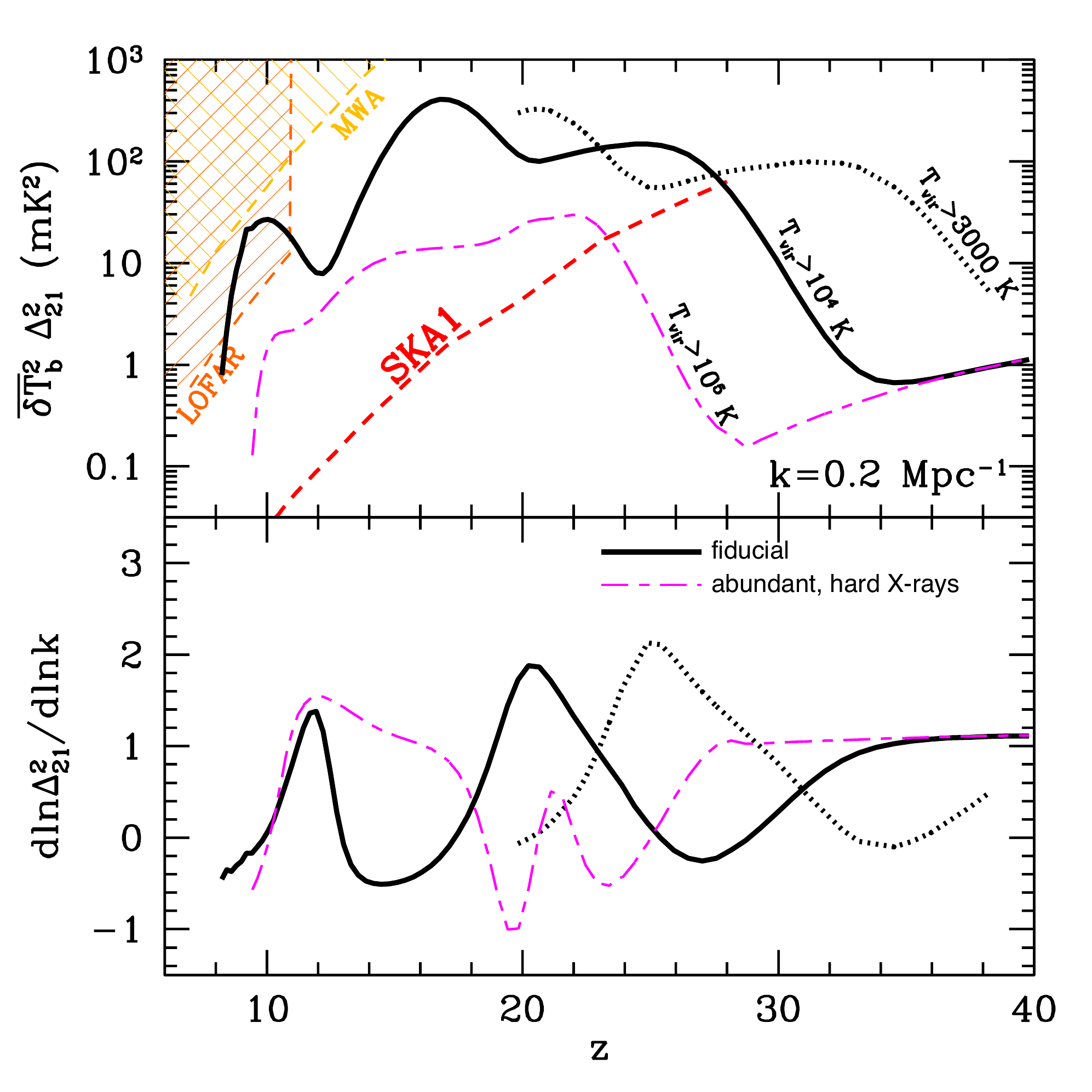}
\includegraphics[width=0.45\textwidth]{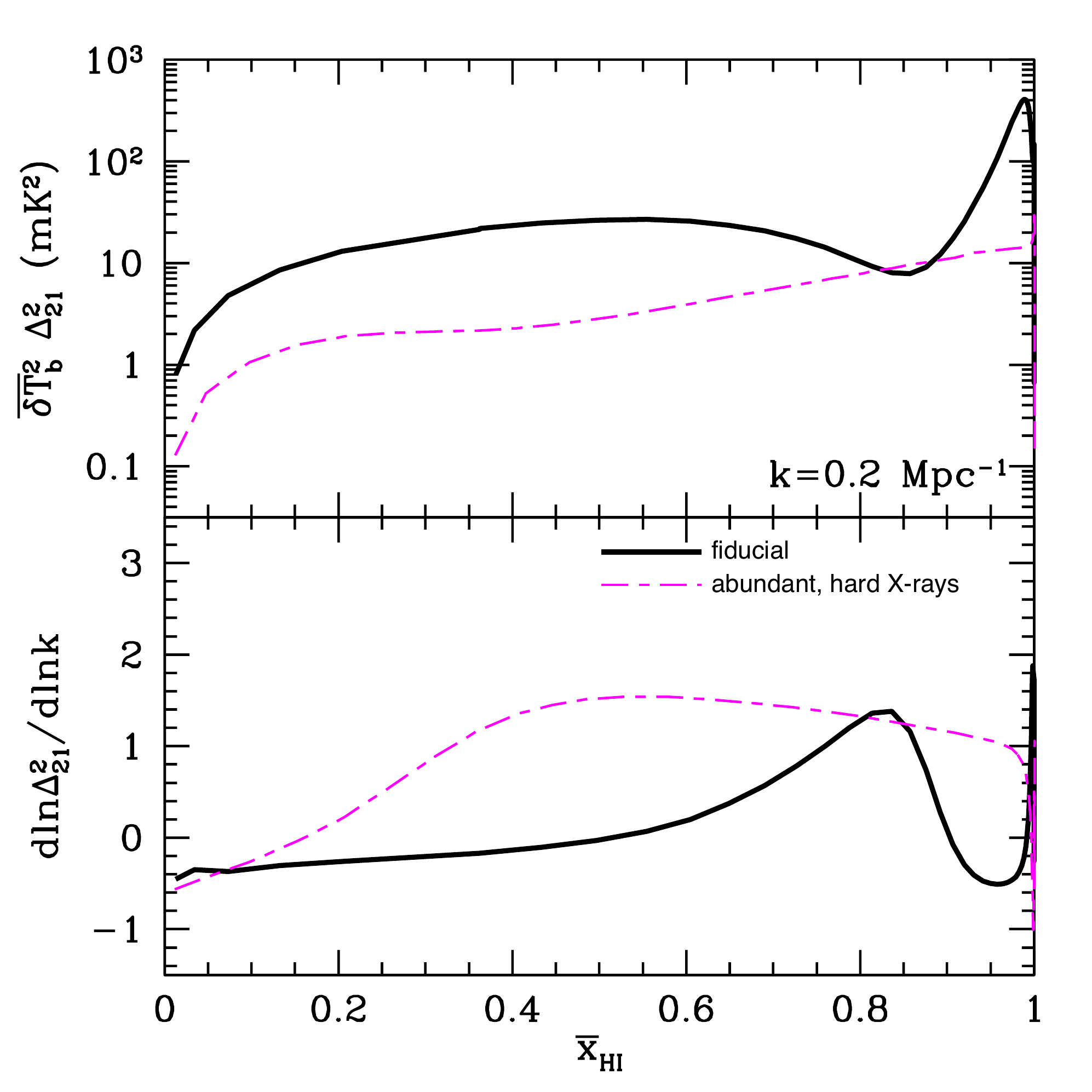}
}
\vspace{-\baselineskip}
\caption{\footnotesize
Evolution of the 21cm power spectrum amplitude ({\it top panels}) and slope ({\it bottom panels}) at $k=0.2$ Mpc$^{-1}$.  Evolution vs redshift ($\avenf$) is shown on the left (right).  Solid black (dashed magenta) curves correspond to the fiducial (extreme X-ray) model shown in the top (bottom) panel of Fig. 1.  The achievable thermal noise after a 1000h observation with SKA1-LOW is shown with the dashed red curve in the left panel; the analogous noise for the first generation instruments, LOFAR and MWA is marked by the checkered region in the top left corner of the panel. As discussed in the text, fiducial models show a three-peaked evolution in the large-scale 21cm power, corresponding to: (i) WF coupling, (ii) X-ray heating; and (iii) EoR. Models are taken from \cite{MFS13}.  Also shown in the left panel is a model with an additional contribution at $z\geq20$ from H$_2$ cooling halos with virial temperatures $> 3000$ K.
}
\label{fig:ps_evo}
\vspace{-0.5\baselineskip}
\end{center}
\end{figure*}

The first galaxies likely formed at high redshifts, $z>30$ in very low mass halos $M_{\rm halo}=10^{6-7}\Msun$ (e.g. \cite{HTL96, ABN02, BCL02}), sometimes referred to as minihalos.  Since the gas was primordial in composition, accretion and star-formation was governed by H$_2$ cooling.  Due to their shallow potential wells and inefficient cooling, the star formation inside minihalos is strongly susceptible to feedback effects, e.g.: (i) mechanical feedback from SNe explosions; (ii) X-ray heating; (iii) ionizing UV background (UVB); (iv) H$_2$ dissociative radiation.  The later, so-called Lyman-Werner background (LWB) is expected to eventually sterilize star-formation inside minihalos (e.g. \cite{HAR00,RGS01,MBH06}).  In fact, even galaxies hosted by more massive, atomically-cooled halos (which are more resilient to feedback) could be sufficiently-abundant to establish a LWB strong enough (e.g. \cite{ON08}) to sterilize star-formation inside minihalos by $z\sim20$ (e.g. \cite{HF12, Fialkov13}; though see \cite{Ahn12}).

Nevertheless, these fragile first galaxies are likely the ones which start the Cosmic Dawn.  As mentioned above, this beginning is observable in 21cm through the WF coupling epoch, mostly driven by photons just redward of Ly$\beta$ which emerge from galaxies and redshift into \lya\ resonance.  These photons have mean free paths of $\gsim 100$ Mpc, and can efficiently couple the spin temperature to the gas temperature, with 21cm being seen in absorption against the CMB.  21cm fluctuations during this epoch are driven by fluctuations in the strength of this coupling, i.e. the \lya\ background.  Therefore, {\it the timing and duration of the initial rise and fall of the 21cm power tells us about the star-formation inside the very first, molecularly-cooled galaxies}.

This is highlighted by comparing the solid and dotted black curves in the left panels of Fig. \ref{fig:ps_evo}.  The solid black curve corresponds to a fiducial model, in which star formation only occurs in atomically-cooled galaxies.  The dotted black curve assumes an additional contribution at $z\geq20$ from minihalos with virial temperatures $> 3000$ K.  The WF epoch starts earlier, and is more extended, driven by the slower evolution of the halo mass function on these scales.  This delay is likely further extended by the feedback processes mentioned above.  In this model, the peak in power associated with WF coupling would occur too early (at low frequencies) to be detectable with SKA1-low.  However, the trough between the WF coupling and X-ray heating epochs should be detectable.  Hence, if we do {\it not} detect a third peak in power at $z\gsim20$, it would imply that minihalos are driving WF coupling.

Moreover, the {\it shape} of the power-spectrum at a given astrophysical milestone can tell us which halos hosted the first galaxies (e.g. \cite{Santos11}).  As we shall also see below, radiation fields driven by more biased, rare galaxies result in less small-scale structure and associated  21cm power.

\section{X-ray properties of early galaxies and IGM heating}
\label{sec:Xrays}

As discussed above, X-rays from early galaxies are expected to heat the IGM to temperatures above the CMB well before the bulk of reionization. When the ionized fraction of the IGM surpasses a few percent, most of the X-ray energy gets deposited as heat through free-free interactions of the primary ionized electron (e.g. \cite{SvS85, FS10, VEF11}).  This makes X-rays more efficient as sources of heating than ionization.  However, they could still significantly contribute to the EoR (in addition to the heating epoch), since the first, metal-poor galaxies are likely more X-ray luminous than local ones (per unit star formation rate; e.g. \cite{Fragos12, Basu-Zych13}.  Current limits do not rule out a strong redshift evolution in this efficiency of X-ray production (e.g. \cite{BKP14}).
X-rays could also indirectly impact the EoR by raising the Jeans mass in the IGM.  The resulting photo-heating feedback on sources might delay the EoR (e.g. \cite{RO04, MFS13}), providing a window for a clean measurement of the matter power spectrum.

The interaction of X-rays with the IGM is characterized by a very large mean free path:
\begin{equation}
\label{eq:mfp}
\lambda_{\rm X} \approx 20 ~ \avenf^{-1} \left( \frac{E_{\rm X}}{\rm 300 eV} \right)^{2.6} \left( \frac{1+z}{10} \right)^{-2} ~ {\rm cMpc} ~,
\end{equation}
where $\avenf$ is the (volume) mean neutral fraction of the IGM and $E_{\rm X}$ is the photon energy.  This means that only soft X-rays ($E_{\rm X} \lsim$ keV) interact with the IGM, making them relevant for 21cm studies.  Without sub-keV X-rays, the EoR and IGM heating epochs would have dramatically different (smoother) morphologies (see the bottom panel of Fig. \ref{fig:delT}).

\begin{figure*}
\vspace{-1\baselineskip}
\begin{center}
{
\includegraphics[width=0.45\textwidth]{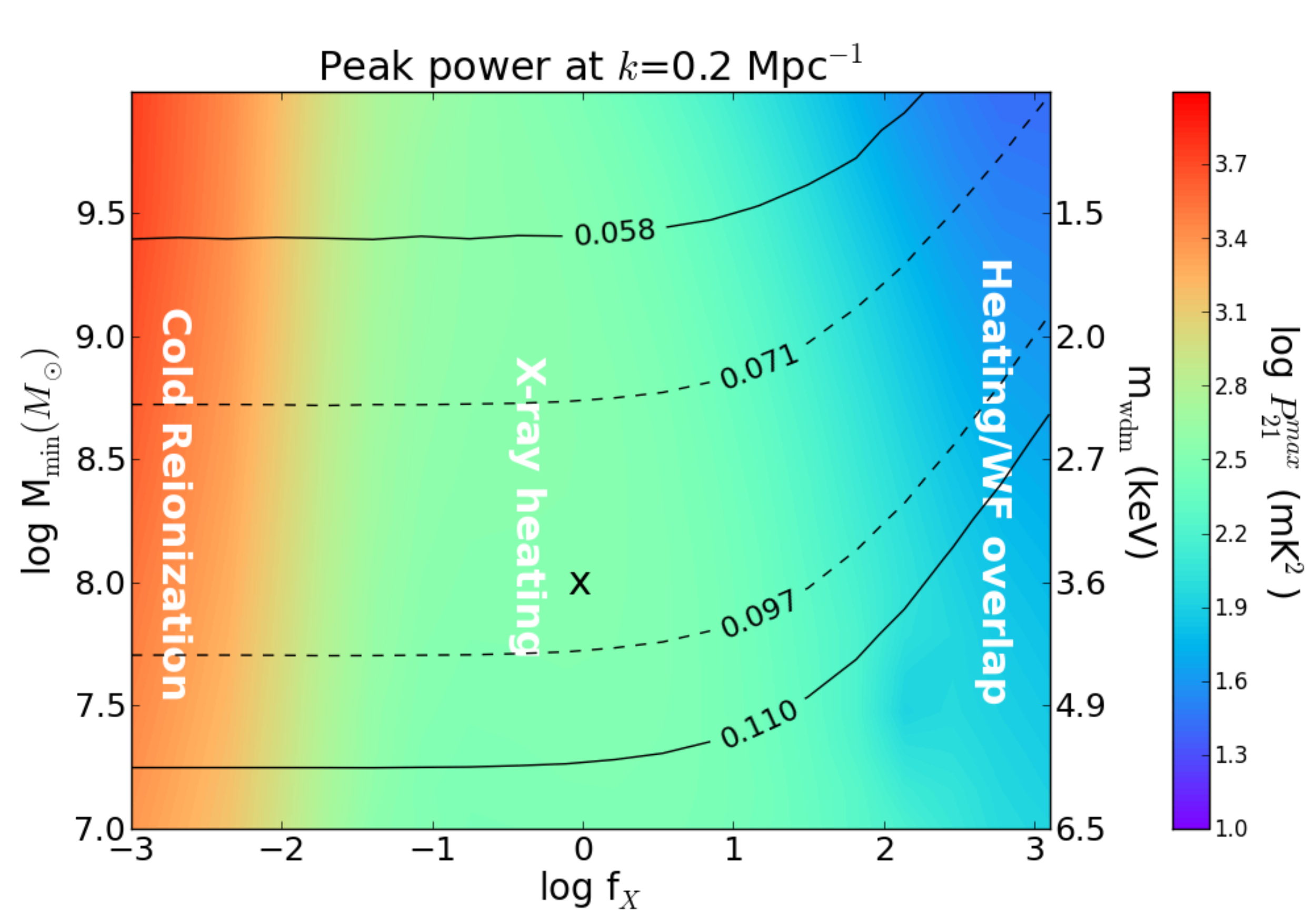}
\includegraphics[width=0.45\textwidth]{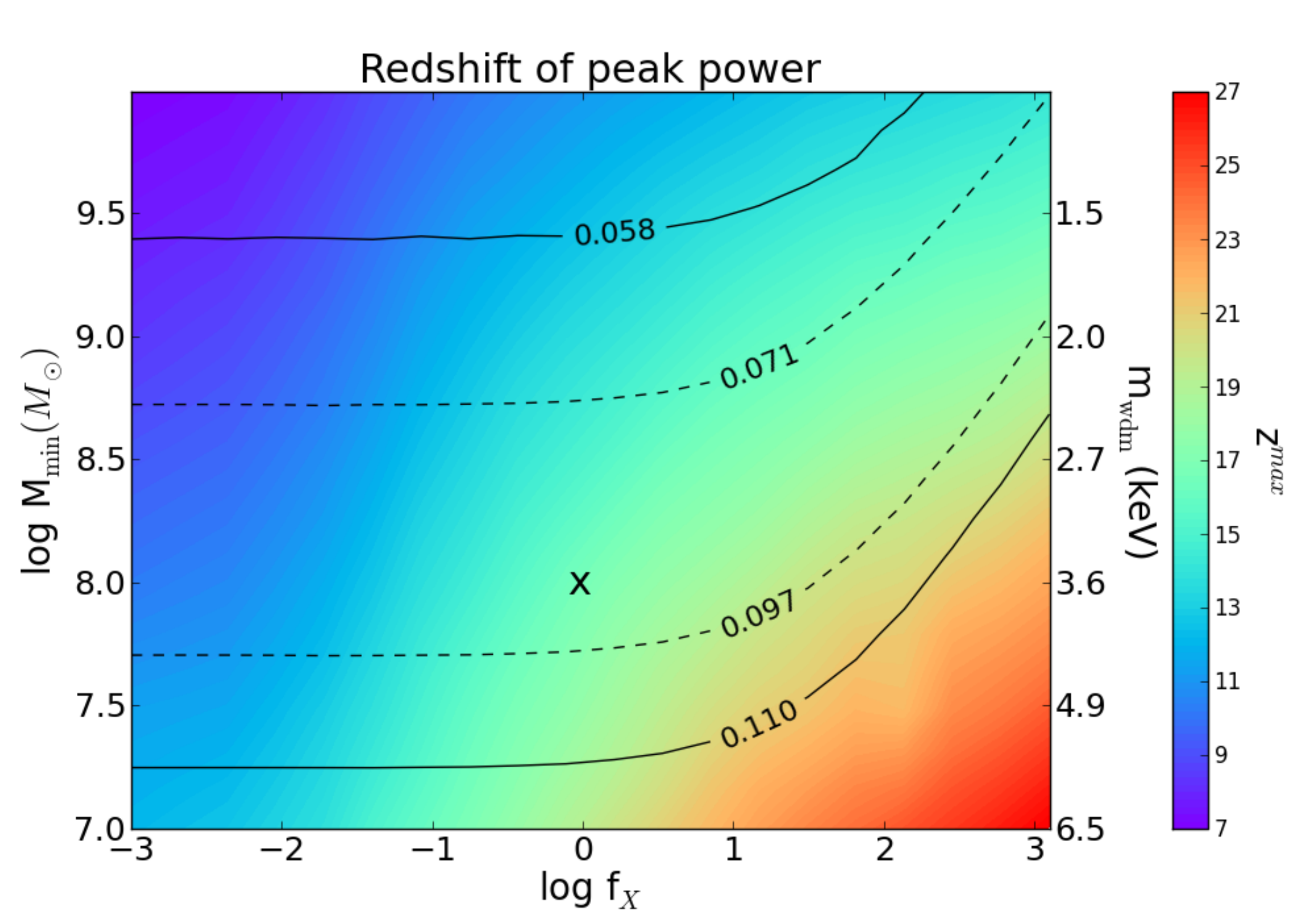}
\includegraphics[width=0.45\textwidth]{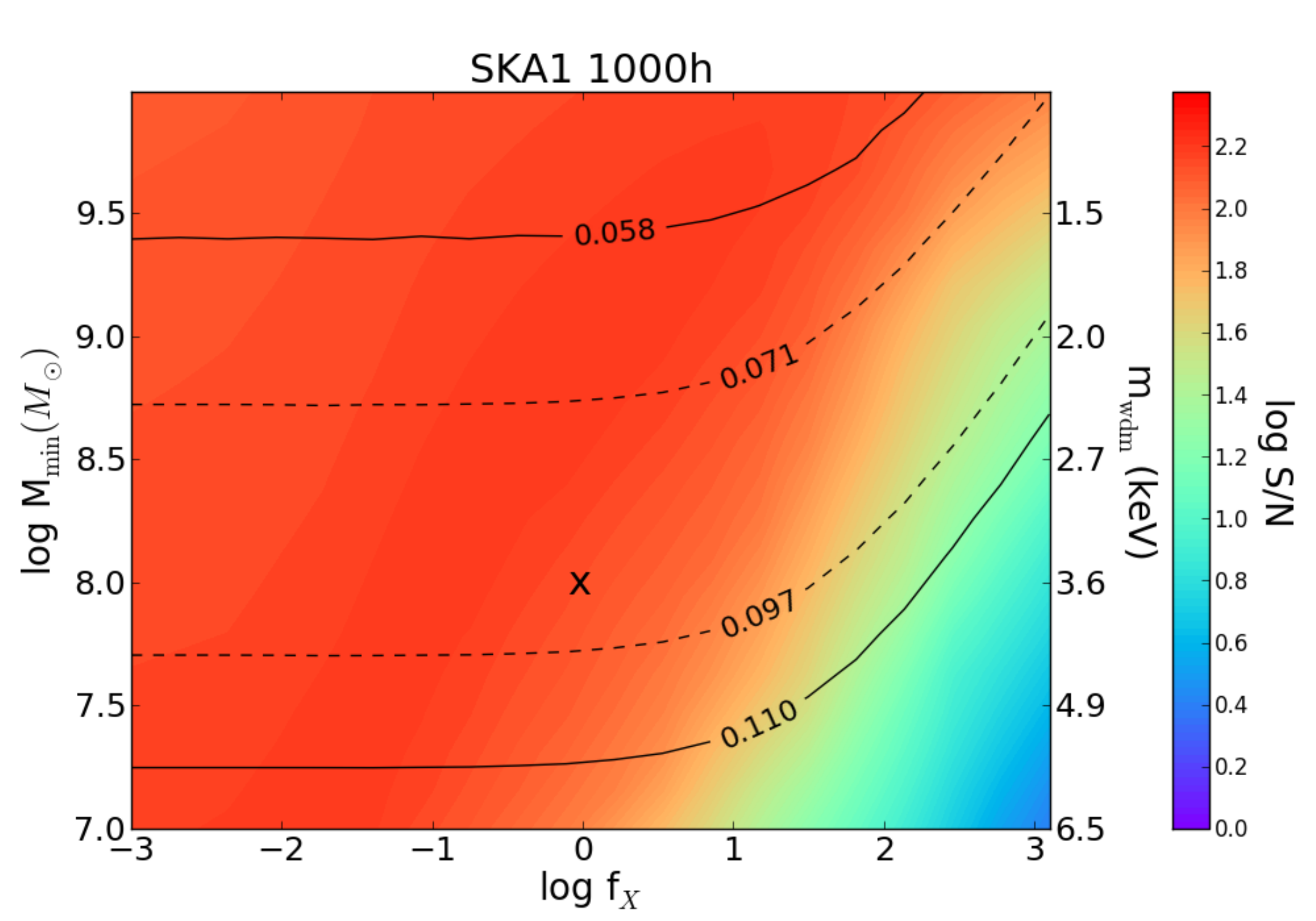}
\includegraphics[width=0.45\textwidth]{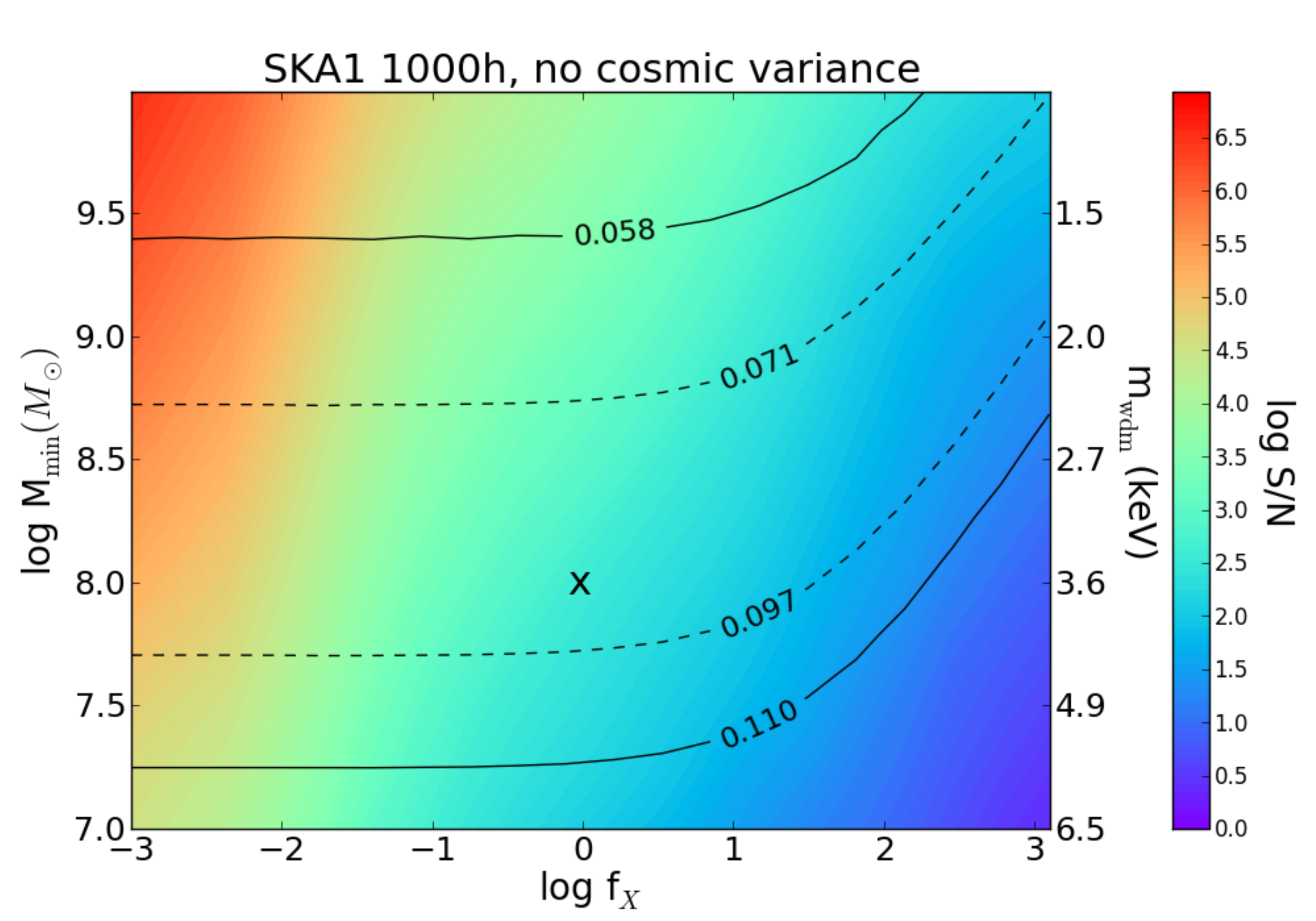}
}
\vspace{-\baselineskip}
\caption{\footnotesize
Signal and S/N evaluated at the redshift when the $k=0.2$ Mpc$^{-1}$ power is the largest, as a function of the minimum halo mass hosting star forming galaxies, $\Mmin$, and their X-ray efficiency, $f_X$. {\it Top left:} peak signal, i.e. maximum amplitude of the 21cm power. {\it Top right:} redshift of the peak signal.  {\it Bottom left:} corresponding S/N after 1000h with SKA1-LOW Phase 1.  {\it Bottom right:} S/N neglecting the cosmic variance component: $P_{21}/\sqrt N_k$.
A 'fiducial' model with $\Mmin=10^8\Msun$ and $f_X=1$, is denoted with an 'x' in the panels.
 For the observing strategies used, see \cite{ME-WH14}.
}
\label{fig:heating}
\vspace{-0.5\baselineskip}
\end{center}
\end{figure*}

IGM heating during the CD is the most potent probe of early X-rays.  X-rays from galaxies likely dominate the heating.  First generation instruments will not detect this exciting epoch, for most reasonable scenarios \cite{ME-WH14}. The SKA1-LOW is poised to offer us the first glimpses into this exciting epoch, driven by the high-energy processes inside the first galaxies!

To further quantify this, in the top panels of Fig. \ref{fig:heating}, we show the peak amplitude and redshift of the peak amplitude of the large-scale 21cm power.  We plot quantities in the parameter space of: (i) the minimum halo mass hosting star forming galaxies, $\Mmin$, and (ii) their X-ray efficiency, $f_X$.  The fiducial value, $f_X \equiv (N_{\rm X}/0.25) = 1$ corresponds to $N_{\rm X}=0.25$ X-ray photons per stellar baryon, consistent with empirical scaling relations from nearby star-forming galaxies (e.g. \cite{MGS12}).  Similarly, the fiducial choice of $\Mmin\sim10^8\Msun$ corresponds to atomic-cooling galaxies (see \S \ref{sec:EoRsources} below). Increasing $f_X$ shifts the X-ray heating epoch (and associated peak in power) towards higher redshifts, while increasing $\Mmin$ shifts {\it all} astrophysical epochs towards lower redshifts (see top left panel of Fig. \ref{fig:ps_evo}).  Therefore, this 2D parameter space should span most of the variation in the signal.

Over a broad swath of parameter space, the large scale 21cm power during heating (driven by temperature fluctuations) peaks at a value of few hundred mK$^2$ (c.f. \cite{Baek10}), an order of magnitude larger than the peak during the EoR (driven by ionization fluctuations).  
  In the bottom panels of Fig. \ref{fig:heating} we show the signal-to-noise (S/N) at which this signal can be detected with a 1000h observation of SKA1-LOW, with and without cosmic variance (left and right panels, respectively).  {\it We see that the SKA1-LOW should easily detect X-ray heating throughout the parameter space}.  Most scenarios are limited by cosmic variance, suggesting an observational strategy of sampling several independent fields, if higher S/N is desired, e.g. for imaging.

What are the likely sources of X-rays in the early galaxies?  The bolometric X-ray luminosities of local, star-forming galaxies are dominated by bright, high mass X-ray binaries (HMXBs; e.g. \cite{GGS04, Mineo_HMXB}).  However, the soft-bands relevant for X-ray heating, $E_X \lsim 2$ keV, have a comparable contribution from the hot interstellar medium (ISM; \cite{Mineo_ISM}).  Because of the strong dependence of the X-ray mean free path on the photon energy (eq. \ref{eq:mfp}), {\it the X-ray spectral energy distributions (SEDs) of the first galaxies have a strong imprint on the 21cm fluctuations during the CD.}
 In particular, scenarios in which X-ray heating is dominated by HMXBs should result in a factor of $\sim3$ less large-scale 21cm power
 than those dominated by the hot ISM \cite{Pacucci14}.  This difference is easily identifiable with the SKA1-LOW, and importantly, is not degenerate with the galaxies' X-ray luminosities.  Hence, {\it the SKA will be a powerful tool for studying the first galaxies and their high-energy processes.}

\section{Radio loud sources at high-redshifts}
\label{sec:heating}

Before the Universe was heated to temperatures greater than the CMB, the IGM could be visible in absorption along sightlines to high-redshift radio sources.  This is the 21cm analogy of the well-studied \lya\ forest.
Although detecting features in the forest will be challenging even for SKA1-LOW (e.g. \cite{MW12}),  integration time can be reduced if one is only after a statistical measure, such as the increased variance along a sightline (e.g. \cite{Carilli04}).  In any case, detecting the 21cm forest along individual sightlines requires a radio-loud QSO at redshifts before heating (e.g. \cite{MW12, Ciardi13}).
 It is not clear how likely this is, given that the population of bright QSOs declines rapidly towards high-$z$ (e.g. \cite{Willott07, Wilman08}).

Chances can be improved however using the statistical imprint of a larger number of fainter QSOs.  A high-redshift population of radio sources will introduce small-scale power in the 21cm power spectrum pre-heating.  Temperature fluctuations driven by the X-ray emitting galaxies are expected to dominate the large-scale power, while the small-scale power ($k\gsim0.5$ Mpc$^{-1}$) will allow us to constrain the population of radio loud active galactic nuclei at high redshifts \cite{E-W14}.

Dense structures such as damped Lyman alpha systems (DLAs), or even sterilized minihalos (mentioned above), would aid in the detection of the 21cm forest, if they contain enough cold, neutral gas.  The resulting increased absorption features would be strongly imprinted in both individual sightlines (e.g. \cite{FL02, MW12}) as well as statistical detections \cite{E-W14}, allowing us to additionally constrain the abundance of cold gas clumps at high-$z$.

\section{EoR sources}
\label{sec:EoRsources}

\begin{figure*}
\vspace{-1\baselineskip}
\begin{center}
{
\includegraphics[width=0.3\textwidth]{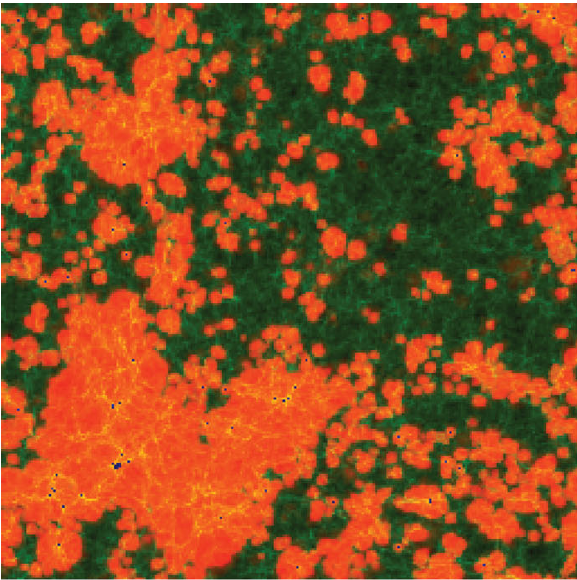}
\includegraphics[width=0.3\textwidth]{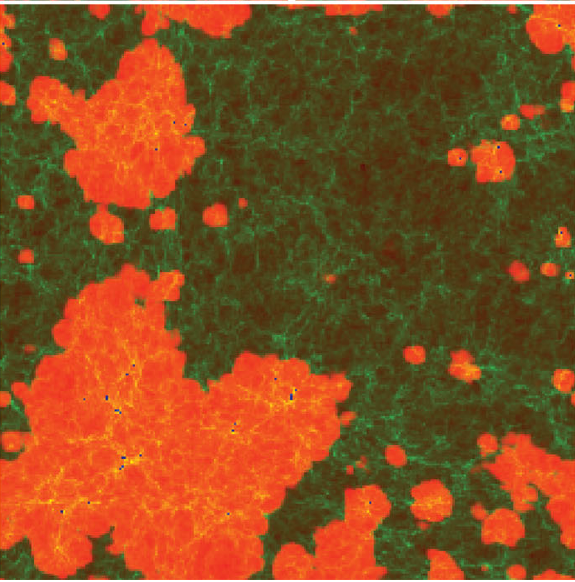}
\includegraphics[width=0.32\textwidth]{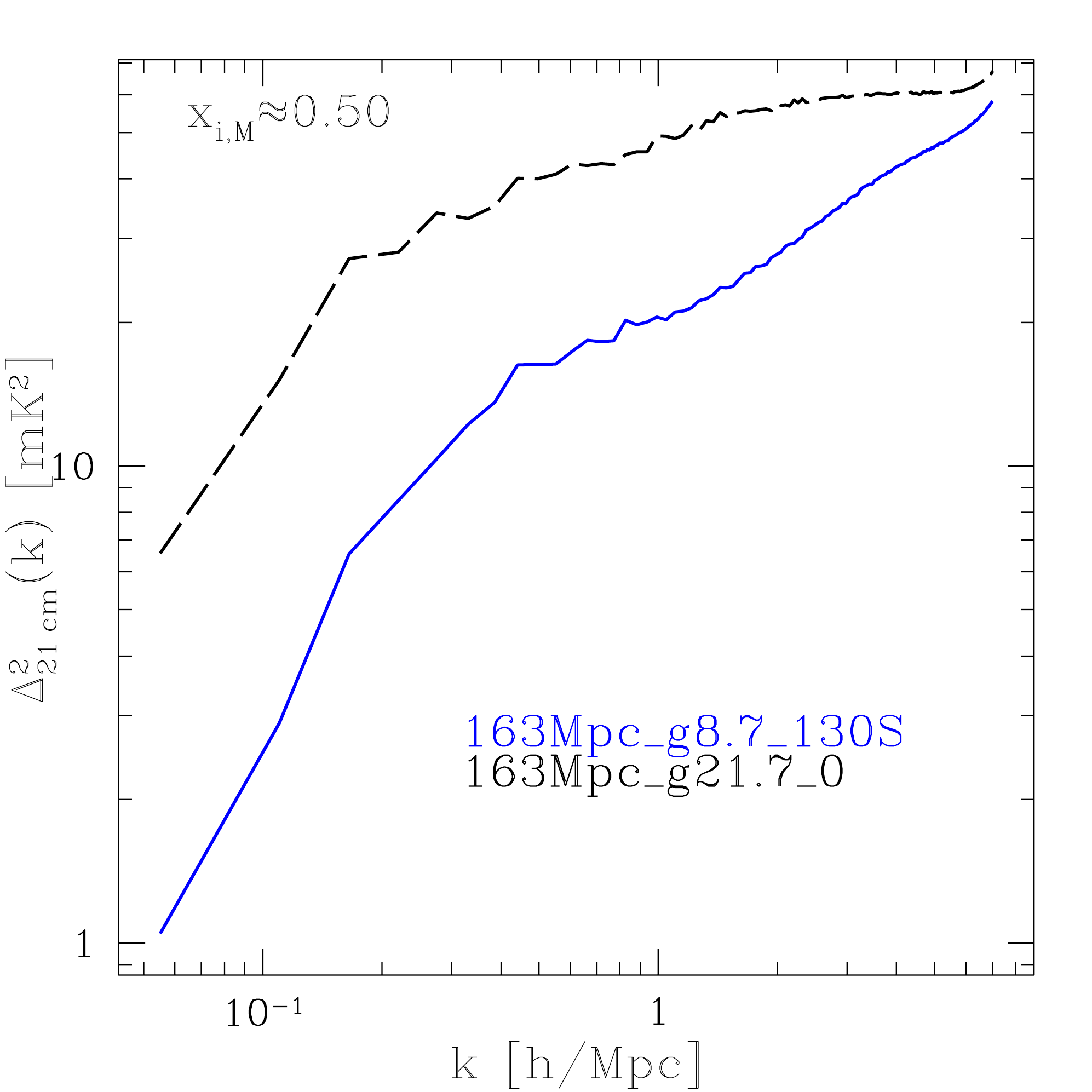}
}
\vspace{-0.5\baselineskip}
\caption{\footnotesize
EoR morphologies driven by bursty star formation in atomically-cooled galaxies, $\Mmin \gsim10^8\Msun$, ({\it left panel}) compared to those driven by galaxies residing in $\Mmin \gsim10^9\Msun$ ({\it middle panel}). Slices are 53 Mpc on a side, and correspond to the mid-point of reionization.  The power spectra in the right panel are generated from 163 Mpc simulations with analogous source prescriptions. The power spectra corresponding to the two source models differ by factors of $\sim$2--3 on large scales.  Panels are from \cite{Iliev12}.
}
\label{fig:EoR_morph}
\vspace{-0.5\baselineskip}
\end{center}
\end{figure*}

The SKA1-LOW we will map out the timing and duration of the EoR (see Fig. \ref{fig:ps_evo}).  The EoR is expected to be driven by UV photons from galaxies (e.g. \cite{McQuinn12}), and is characterized by some generic features in the evolution of 21cm power.
 Initially (following heating), there is a dramatic drop in large-scale amplitude and steepening of the slope during the early stages of reionization, $\avenf\gsim0.8$, when the burgeoning HII regions 'cover-up' the densest IGM patches  (solid curve in the right panel of Fig. \ref{fig:ps_evo}).  This evolution can be understood (to first order, assuming $T_S\gg\Tcmb$) as the transition of the 21cm power spectrum, $P_{21} \approx P_{xx} - 2\avenf P_{\rm xd} + \avenf^2 P_{\rm dd}$, from being dominated by the density power, $P_{dd}$,  to the ionization power, $P_{xx}$; the transition being governed by the negative contribution of the density-ionization cross spectrum, $P_{xd}$   \cite{Lidz08}.  Subsequently, the fluctuations in the ionization field, $P_{xx}$, drive the large-scale 21cm power to a peak value during the mid-point of reionization.

However, this evolution is qualitatively different if hard ($\gsim1$ keV) X-rays play a dominant role.  The resulting weaker density-ionization cross-power and ionization power results in a much more gradual fall in power during the EoR, as well as a much lower overall amplitude (dashed curve in the right panel of Fig. \ref{fig:ps_evo}).  Both of these scenarios are easily identifiable with SKA1-LOW.  More reasonable models, in which (soft) X-rays play a sub-dominant role in the EoR still show a 21cm peak in large-scale power during the midpoint of reionization.  However, the $P_{xd}$ induced drop in power occurs earlier, $\avenf\sim0.9$, due to the pre-ionization from X-ray sources \cite{MFS13}.  The 21cm probability distribution function (PDF) can also be used as a diagnostic, with X-rays decreasing the bi-modality of the PDF \cite{Baek10}.


 In addition to the timing and duration of the EoR, the 21cm signal can tell us about the nature of EoR galaxies.  The EoR is likely driven by faint galaxies below the sensitivity limits of current and upcoming space telescopes, including {\it JWST}. 
 The EoR galaxies formed through atomic cooling, requiring host halo virial temperatures of $\Tvir\gsim10^4$ K ($\gsim10^8\Msun$ at $z\sim10$).  However, the efficiency of star formation inside these dwarf galaxies at high-$z$ is highly uncertain.  Moreover, it is likely that feedback processes (either through internal mechanical feedback, e.g. \cite{SH03}, or photo-heating feedback from the UVB; e.g. \cite{TW96}), regulated the evolution of their star formation rate (SFR). 
 The timing of the EoR signal, tells us when the dominant EoR sources appeared, which is a combination of their host halo masses and star formation efficiencies.

 This degeneracy can be broken using the EoR morphology, which allows us to study the dominant host halo population of EoR galaxies.
  More massive host halos are more biased, with corresponding EoR morphologies with less small-scale ionization structure (see Fig. \ref{fig:EoR_morph}).  This observation is however complicated by the fact that the bias of the halos evolves only weakly over this mass range \cite{McQuinn07}, and by the fact that sinks of ionizing photons can also strongly impact the EoR morphology (see next section).


\section{EoR sinks}
\label{sec:EoRsinks}

\begin{figure*}
\vspace{-1\baselineskip}
\begin{center}
{
\includegraphics[width=0.32\textwidth]{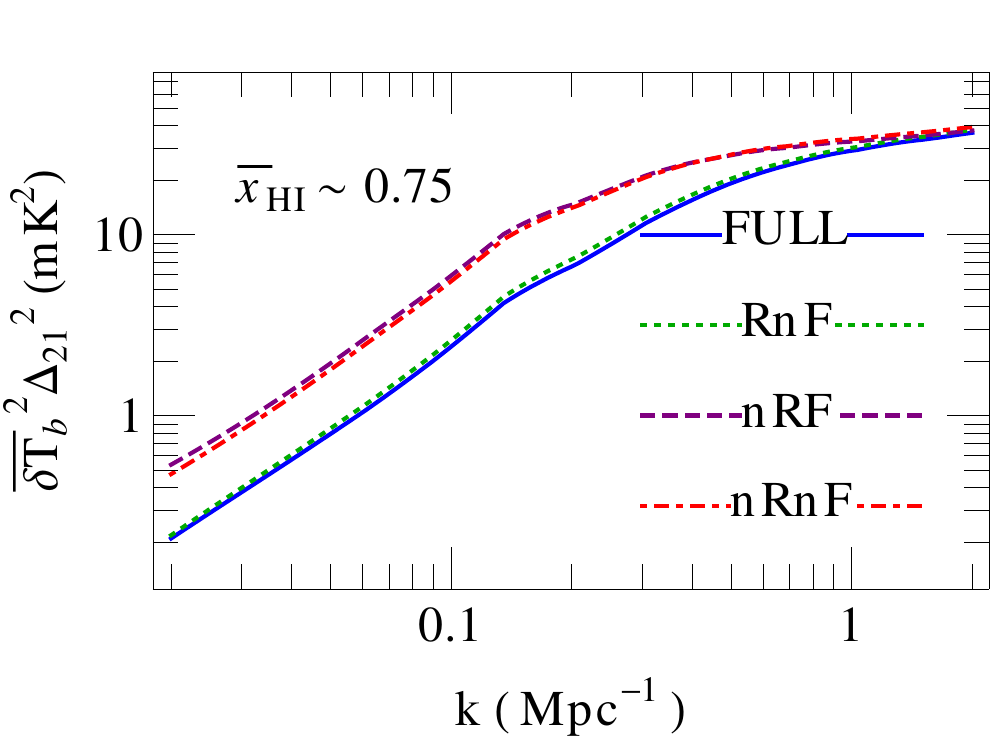}
\includegraphics[width=0.32\textwidth]{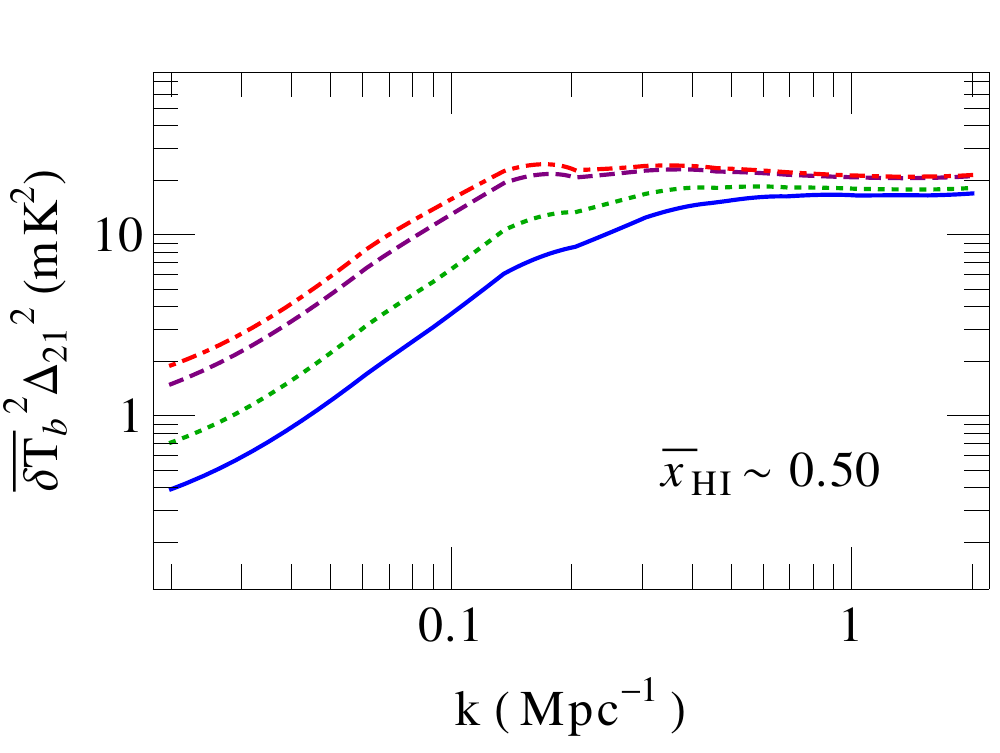}
\includegraphics[width=0.32\textwidth]{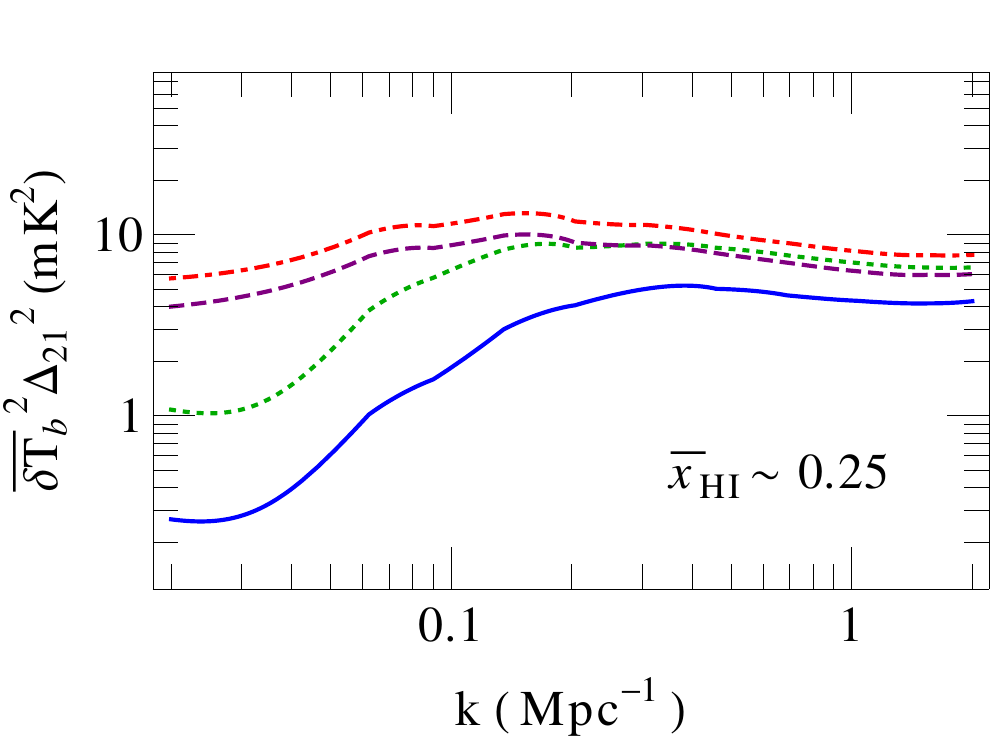}
}
\vspace{-\baselineskip}
\caption{\footnotesize
How 21cm power spectra are affected by {\it inhomogeneous, sub-grid} (i) UV photo-heating feedback on sources and (ii) recombinations in sinks.
The FULL ({\it blue curves}) model includes both effects via calibrated, sub-grid models.
The RnF ({\it green curves}) model includes only recombinations.
The nRF ({\it purple curves}) model includes only UVB feedback.
The nRnF ({\it red curves}) model includes neither effects, and is comparable to current large-scale RT simulations of reionization (e.g. \cite{Zahn11, Majumdar14}).  
 Recombinations in sub-grid structures (and UVB photo-heating feedback to a lesser extent) can strongly suppress large-scale ionization structure (by factors of 2--5 throughout reionization).  Figures are taken from \cite{SM14}.
}
\label{fig:ps_feedback}
\vspace{-0.5\baselineskip}
\end{center}
\end{figure*}

By depleting the ionizing photon budget available to expand cosmic HII regions, recombining systems (largely comprised of so-called Lyman limit systems; LLSs) can have a large impact during (and following) cosmic reionization.   Unfortunately, directly resolving such small ($\lsim$kpc) sinks of ionizing photons in large-scale reionization simulations is computationally impractical.  Therefore, relatively untested sub-grid models must be used.  SKA1-LOW will allow us to confront the predictions of such sub-grid models, offering insights into the small-scale baryonic physics of the IGM.

During reionization, gas is heated from low temperatures to $\sim10^4$ K. 
Depending on the level of pre-heating by X-rays, this might have a strong impact on the clumpiness of gas in the IGM (e.g. \cite{PSS09, ETA13}) and minihalos (e.g. \cite{SIR04, ISR05}).
Subsequently, the gas establishes photo-ionization equilibrium with the local ionizing background, with most recombinations occurring in systems with densities large enough to partially self-shield (e.g. \cite{MOF11}).  As a cosmological HII region grows, an increasing fraction of its ionizing photon budget is lost to balance recombinations, limiting its further growth \cite{FO05}.  This 'slowdown' is more dramatic for larger HII regions, since (i) they generally have a higher ionizing background, with ionization fronts driven into more dense, rapidly recombining systems; and (ii) their centers were the first to be reionized, thus having had enough time to recombine \cite{SM14}.  As large regions slow down, moderate HII regions can 'catch' up, further resulting in a more uniform size distribution even early in reionization.

As a result, recombinations strongly suppress the large scale ionization structure during the EoR.  This can be seen from Fig. \ref{fig:ps_feedback}: comparing the green and red curves, we see that inhomogeneous recombinations can suppress the large-scale 21cm power by factors of 2--3.  The effect is magnified by UVB feedback, which is also strongest inside large HII regions, whose cores ionized early enough for gas to respond to photo-heating.  Comparing the blue and red curves, we see that the total impact of both (i) recombinations and (ii) UVB photo-heating feedback on sources results in a suppression of large-scale 21cm power by factors of 2--5.  Neither effects are accounted for in most large-scale EoR simulations.

Although 'directly' detecting sinks through a corresponding rise in 21cm power on small-scales has been suggested \cite{CHR09}, realistic models of their sizes and neutral gas mass make this direct detection unlikely even with the SKA1-LOW \cite{SM14}.  However, the indirect detection via the above-mentioned steep slope of the 21cm power spectrum should be easy to detect and can tell us about small-scale IGM structure.

\section{Observing Strategy}
\label{sec:obs}

Different observing strategies can impact the scientific return from SKA1-LOW.  For example, the large-scale modes are dominated by sample (cosmic) variance, 
while the small-scale modes are affected by the intrinsic detector (thermal) noise. 
Increasing the total integration time of a single observed patch of the sky decreases the noise on small-scales, while instead observing many patches decreases the cosmic variance.

A quantitative analysis of EoR/CD constraints with different observing strategies has not been done yet.  Qualitatively, we expect the largest constraining power to come from the redshift evolution of large-scale modes (e.g. \cite{ME-WH14, Pober14}). From figures  \ref{fig:ps_evo} and \ref{fig:heating} we see that a single 1000h observation with SKA1-LOW, even with a possible 50\% reduction in area, will be limited by cosmic variance on large-scales.  This suggests that optimal science returns would benefit from an increase in the total field of view, at the expense of a modest reduction in integration time.   These considerations motivate SKA1-LOW's current planned three-tiered observing strategy: (i) 1x1000h; (ii) 10x100h; (iii) 100x10h.  The relative science gains from these strategies will depend on: (i) the effectiveness of foreground removal on large-scales; (ii) statistics used to characterize the cosmic signal; as well as (iii) the available instantaneous bandwidth probing the redshift evolution of the signal.


\section{Conclusions}
\label{sec:conc}

We summarize our main conclusions below.

\begin{packed_item}

\item The timing and duration of the initial rise and fall of the 21cm power tell us about star-formation inside the very first, molecularly-cooled galaxies.  Efficient star-formation inside these minihalos could source WF coupling fluctuations (driven by the \lya\ background) as early as $z\gsim30$.  Hence if we do {\it not} detect a third peak in 21cm power on large scales at $z\lsim30$ with SKA1-LOW, it would imply that minihalos are driving WF coupling.

\item The peak {\it amplitude} of the large-scale 21cm power during X-ray heating is sensitive to the SEDs of the first galaxies (at the factor of few level), while the {\it redshift} of the peak tells us about their X-ray luminosity and host DM halos.

\item The early stages of reionization, $\avenf\approx$0.8-0.9, are characterized by a drop in large-scale 21cm power amplitude, and a steepening of the power spectrum.  The timing and duration of this feature tells us about the contribution of X-rays to the EoR.

\item The midpoint of the EoR is characterized by a local maximum in large-scale 21cm fluctuations (driven by the ionization field), except if hard X-rays dominate reionization.

\item The morphology of the EoR encodes information about the efficiency of star-formation inside galaxies, feedback processes and absorption systems.


\item Absorption systems inhibit the growth of large HII regions, and could result in a dramatic reduction of large-scale 21cm power (by factors of 2--5).  Therefore the steepness of the EoR power spectrum can tell us about the structure and evolution of these small-scale gas clumps.

\item Radiation fields driven by more biased, rare galaxies result in less small-scale structure.  Thus the 21cm power on small-scales encodes information on the host halos which host the dominant sources.

\item The 21cm forest, if detected either through individual sightlines or statistically through the rise in 21cm power towards small-scales ($k\gsim0.5$ Mpc$^{-1}$), will constrain the population of radio-loud AGN as well as cold gas clumps at high-$z$.

\end{packed_item}

\bibliographystyle{unsrt}
\bibliography{ms}{}

\end{document}